\documentclass[aps,prl,twocolumn,superscriptaddress,showpacs]{revtex4}

\usepackage{graphicx}
\usepackage{amsmath}

\begin{document}

\title{Ring--Shaped Andreev Billiards in  Quantizing Magnetic Fields}

\author{J. Cserti}
\email{cserti@galahad.elte.hu}
\author{P. Polin\'ak}
\author{G. Palla}
\affiliation{Department of Physics of Complex Systems, E{\"o}tv{\"o}s
University, H-1117 Budapest, P\'azm\'any P{\'e}ter s{\'e}t\'any 1/A,
Hungary}

\author{U. Z\"ulicke}
\email{u.zuelicke@mailaps.org}
\affiliation{Institut f\"ur Theoretische Festk{\"o}rperphysik,
Universit{\"a}t Karlsruhe, D-76128 Karlsruhe, Germany }

\author{C.~J. Lambert}
\email{c.lambert@lancaster.ac.uk}
\affiliation{Department of Physics, Lancaster University, Lancaster,
LA1 4YB, UK}

\date{\today}

\begin{abstract}
We present a detailed semiclassical study of a clean disk--shaped
insulator--normal-metal--superconductor hybrid system in a magnetic
field. It is based on an exact secular equation that we derived within
the microscopic Bogoliubov--de~Gennes (BdG) formalism. Results
obtained from a classification of electron and hole orbits are in
excellent agreement with those from an exact numerical diagonalization
of the BdG equation. Our analysis opens up new possibilities for
determining thermodynamic properties of mesoscopic hybrid systems.
\end{abstract}


\maketitle

Mesoscopic hybrid systems consisting of normal metals (N) in contact
with superconductors (S) exhibit interesting and sometimes
counterintuitive equilibrium and transport properties resulting from
the interplay between quantum--mechanical phase coherence and
superconducting correlations~\cite{hekkingreview,lambertreview}. A
prominent example is the paramagnetic re-entrance effect observed
recently in experiments performed by Visani et al.~\cite{Visani} on
cylindrical S--N proximity samples. While numerous theoretical
works~\cite{Bruder,ellenzek,disord,andrei} have addressed this
problem, a fully satisfactory explanation of the origin of the sizable
paramagnetic contribution to the susceptibility is still lacking.
These previous works have studied the spectrum of Andreev bound states
formed in planar normal--metal layers in contact with a bulk
superconductor, neglecting the effects of cyclotron motion of
electrons and holes due to the external magnetic field. Our work
presented here extends these studies, taking into account the
experimentally relevant circular geometry and fully accounting for
quantum effects due to the applied magnetic field. Solving  exactly
the microscopic Bogoliubov-de Gennes  equation (BdG)\cite{BdG-eq}, the
Andreev levels in a cylindrical NS system are obtained for arbitrary
magnetic field. In addition, we give a complete semiclassical
description of the spectrum by identifying the possible classical
orbits corresponding to the quantum states. This analysis is based on
methods developed in our previous work~\cite{box_disk:cikk} which have
been adapted to the case of a finite magnetic field following
Ref.~\onlinecite{Ulrich}. Besides being useful for shedding further
light on causes for the above--mentioned paramagnetic re-entrance
effect, our results are intended to serve as a stimulating guide to
the investigation of proximity effects in large magnetic fields, which
was the focus of several recent experimental~\cite{strongBexp} and
theoretical~\cite{strongBtheo} works.

We consider a superconducting disk of radius $R_S$ surrounded by a
normal metal region of radius $R_N$. (This models the experimentally
realized~\cite{Visani} cylindrical geometry because motion along the
axis of the cylinder adds only a trivial kinetic--energy term.) 
The magnetic field is perpendicular to the plane of the disk with 
a constant value of $B$ in the N region and zero inside the S region. 
Thus, the non-zero component of the vector potential 
in polar coordinates $(r,\vartheta)$ with symmetric gauge is 
given by \cite{Solimany} $ A_\vartheta(r,\vartheta) = B\, (r^2 -R_S^2)
\, \Theta(r-R_S)/(2r^2)$, where $\Theta(x)$ is the Heaviside function.
Excitations in an NS system are described by the BdG equation:
\begin{equation}
\left(\begin{array}{cc}
H_0  & \Delta \\ 
\Delta^* & -H_0^* 
\end{array}   
 \right)\Psi = E \,  \Psi,
\end{equation}
where  $\Psi$ is a two-component wave function, and $H_0 = {\left(
{\bf p} - e{\bf A}\right)}^2/(2m_{\rm eff})+ V-E_{\rm F}$. Fermi
energies and effective masses in the S and N region are denoted by 
$E_{\rm F}= E_{\rm F}^{\left(\rm S\right)}, \,\, E_{\rm F}^{\left(\rm 
N\right)}$ and $m_{\rm eff} = m_{\rm S}, \,\,m_{\rm N}$, respectively.
$e$ is the electron charge. In the limit $R_N-R_S \gg \xi_0$, the
superconducting pair potential can be approximated by a step function 
$\Delta({\bf r})=\Delta_0\Theta(R_s-r)$, where $\xi_0 = \hbar v_{\rm
F}/\Delta_0$ is the coherence length, and $v_{\rm F}$ is the Fermi
velocity. Self-consistency of the pair potential is not taken into
account, similar to the treatment given in Ref.~\onlinecite{BTK}. At
$r=R_N$, Dirichlet boundary conditions (i.e., an infinite potential
barrier) are assumed, while at the NS interface $r=R_S$, the presence
of a tunnel barrier is modeled by a delta potential $V(r)=U_0\, \delta
(r-R_S)$. The energy levels of the system are the positive eigenvalues
$E$ of the BdG equation. In what follows, we consider the energy
spectrum below the superconducting gap, $0 < E < \Delta_0$. Rotational
symmetry of the system implies a separation ansatz for the wave
function as a product of radial and angular parts. We choose for the
angular part the appropriate angular--momentum eigenfunctions with
quantum number $m$. Then the radial wave functions $f^{\pm}_m(r)$
satisfy a one-dimensional BdG eq.\ in the normal region:
\begin{equation}
h_0^{\left(\pm\right)} f^{\pm}_m(r) =  \varepsilon f^{\pm}_m(r), 
\label{BdG-rad:eq}
\end{equation}
where  $h_0^{\left(\pm\right)}  = -\frac{\hbar \omega_c}{2}\left[2
\frac{d}{d \xi}\left(\xi \frac{d}{d \xi}\right)-\frac{m_{\pm}^2}{2
\xi} -\frac{\xi}{2} + m_{\pm} +\nu_0 \right]$ with new dimensionless
variables $\xi=r^2/(2l^2)$ and $\varepsilon = E/(\hbar \omega_c)$.
The functions $f^{\pm}_m(r)$ are, respectively, the electron and hole
components of the radial Bogoliubov-de Gennes spinor. Here $\omega_c =
|eB|/m_{\rm eff}$ is the cyclotron frequency, $l=\sqrt{\hbar c/|eB|}$
the magnetic length, $\nu_0=2E_{\rm F}^{\left(\rm N\right)}/(\hbar
\omega_c)$, $m_{\pm} = \xi_S \pm m\,\rm{sgn}(eB)$, $\xi_S = R_{\rm
S}^2/(2l^2)$, and $\rm{sgn}(x)$ denotes  the sign function. After
transforming the wave functions $f^{\pm}_m(\xi)\rightarrow\xi^{m_\pm/
2} \, e^{-\xi/2}\, f^{\pm}_m(\xi)$, Eq.~(\ref{BdG-rad:eq}) results in
a Kummer differential equation~\cite{Abramowitz}, and  the ansatz for
the wave function in the normal region ($R_S <r < R_N$) can finally be
written as
\begin{subequations} 
\begin{eqnarray}
\Psi_N (r,\vartheta)\! \! \!  &=& \! \! \! 
\left(\begin{array}{c} a_+ \, \varphi^{({\rm N},+)}_m (r) \\ 
a_- \, \varphi^{({\rm N},-)}_m (r)
\end{array}
\right) e^{im\vartheta} , \,\,\, 
\label{hullfv-N}  \\
\varphi^{({\rm N},\pm)}_m (r) \! \! \! &=& \! \! \! 
\xi^{m_\pm/2}\,e^{-\xi/2}
\left[
M\left( \frac{1}{2}\mp\varepsilon-\frac{\nu_0}{2},1+m_\pm,\xi\right)
\right.  
\nonumber\\
& &  \left.   \hspace*{-1cm} -
{\left(\frac{\xi_N}{\xi}\right)}^{m_\pm}
M\left(\frac{1}{2}\mp\varepsilon-\frac{\nu_0}{2}-m_\pm,1-m_\pm,\xi\right)
\right.
\nonumber\\
& &  \left.   
\hspace*{-1cm} \times 
\frac{M\left( \frac{1}{2}\mp\varepsilon-\frac{\nu_0}{2},1+m_\pm,\xi_N \right)}
{M\left(\frac{1}{2}\mp\varepsilon-\frac{\nu_0}{2}-m_\pm,1-m_\pm,\xi_N\right)}
\right],
\end{eqnarray}
\end{subequations}
where $M(a,b,x)$ is Kummer's function~\cite{Abramowitz}, and $\xi_N=
R_N^2/(2l^2)$. These wave functions satisfy the Dirichlet boundary
conditions at $r=~R_{\rm N}$, i.e., $\varphi^{({\rm N},\pm)}_m (R_{\rm
N})=0$, and the following symmetries hold: $\varphi^{({\rm N},-)}_m(r,
\varepsilon,B)=\varphi^{({\rm N},+)}_m (r,-\varepsilon,-B)$ and
$\varphi^{({\rm N},\pm)}_m (r,\varepsilon,-B)=\varphi^{({\rm N},
\pm)}_{-m} (r,\varepsilon,B)$, where the dependencies on $\varepsilon$
and $B$ are emphasized for clarity.

In the superconducting region $r < R_S$, the ansatz for BdG wave
functions is given by~\cite{box_disk:cikk}: 
\begin{equation}
\Psi_S (r,\vartheta) =\! \! \left[
\! c_+ \! \! \left(\! \!\begin{array}{c} \gamma_+ \\ 
1 \end{array} \! \! \right)\! \varphi^{({\rm S},+)}_m (r) 
+ \! c_- \! \! \left( \! \!\begin{array}{c} \gamma_- \\ 
1 \end{array}\! \! \right) \! \varphi^{({\rm S},-)}_m (r) 
\! \right]  e^{im\vartheta},
\label{hullfv-S}
\end{equation}
where $\varphi^{({\rm S},\pm)}_m (r) = J_m(q_\pm \, r)$, $q_\pm =
k_{\rm F}^{({\rm S})} \, \sqrt{1\pm  i\eta}$, $\eta = \sqrt{\Delta_0^2
- E^2}/E_{\rm{F}}^{\rm{(S)}}$, $\gamma_\pm  = \Delta_0/(E \mp  i
\sqrt{\Delta_0^2 - E^2})$, and $J_m(r)$ is a Bessel function of order
$m$. These satisfy the symmetries $\varphi^{({\rm S},-)}_m (r,
\varepsilon)=\left[\varphi^{({\rm S},+)}_m (r,-\varepsilon)\right]^*$ and $\gamma_-=\gamma_+^*$.

The four coefficients $a_{\pm},  c_{\pm}$ in Eqs.~(\ref{hullfv-N}) and
(\ref{hullfv-S}) are determined from matching conditions at the
interface of the NS system~\cite{box_disk:cikk}. These yield a secular
equation for the eigenvalues $\varepsilon$  of the NS system for fixed
mode index $m$. Using the fact that the wave functions $\varphi_m^{(N,
+)}$ given in Eq.\ (\ref{hullfv-N}) are real functions and the
symmetry relations between the electronic and hole-like component of
the BdG eigenspinor, the secular equation can be reduced to
\begin{subequations}
\begin{equation} 
\rm{Im} \left \{\gamma_+ D^{\rm{(+)}}_m(\varepsilon,B) \, 
D^{\rm{(-)}}_m(\varepsilon,B) \right \}=0,
\label{DNS}
\end{equation}
where 
\begin{equation}
D^{\rm{(+)}}_m(\varepsilon,B) \!=\!
\left| \!\! \begin{array}{cc}
 \varphi_m^{(N,+)} & \varphi_m^{(S,+)}  \\
{\left[\varphi_m^{(N,+)}\right]}^\prime & 
Z\varphi_m^{(S,+)}\! + \!\frac{m_{\rm{N}}}{m_{\rm{S}}} 
{\left[\varphi_m^{(S,+)}\right]}^\prime
\end{array}  \!\! \right| 
\label{De}
\end{equation}
\end{subequations}
and $D^{\rm{(-)}}_m(\varepsilon,B)={\left[D^{\rm{(+)}}_m(-\varepsilon,
-B)\right]}^* $. Here  $Z= \left(2m_{\rm{N}}/\hbar^2\right) \, U_0$ is
the normalized barrier strength, and the prime denotes the derivative
with respect to $r$. All functions are evaluated at $r=R_{\rm S}$. The
energy levels of the NS systems can be found by solving the secular
equation (\ref{DNS}) for $\varepsilon$ at a given quantum number $m$.
The secular equation derived above is exact in the sense that the
usual Andreev approximation is not assumed~\cite{Colin-review}. An
analogous result was found previously~\cite{box_disk:cikk} for zero
magnetic field where the wave functions $\varphi_m^{(N,\pm)}$ are
different.

We now turn to the semiclassical treatment of the system. For
simplicity, we assume a perfect NS interface, i.e., $Z=0$,
$E_{\rm F}^{\left(\rm S \right)}= E_{\rm F}^{\left(\rm N\right)}$,
$m_{\rm S}= m_{\rm N}$. We follow the method developed in 
Ref.~\onlinecite{box_disk:cikk}, i.e., wave functions  which appear 
in Eq.~(\ref{De}) are approximated semiclassically. To construct these
wave functions in the N region, one can use the standard WKB
technique (see, e.g., Refs.~\onlinecite{Klaus} and \onlinecite{Klama})
for the radial Schr{\"o}dinger equation with radial potential given by
$V_m^{\left(\pm\right)}(\tau)={\left(\tau^2/2 -m_{\pm}\right)}^2/
\tau^2$. Here $\tau = r/l$. The four turning points (two each for the
electron and the hole) can be obtained from $V_m^{\left(\pm\right)}
(\tau)=\varepsilon$. This yields
\begin{equation}\label{turnpoints}
\tau_{1,2}^{\pm}=\sqrt{2(\nu_{\pm}+m_{\pm})\mp 2\sqrt{\nu_{\pm}
(\nu_{\pm}+2m_{\pm})}} \quad ,
\end{equation}
where the sign in front of the second term under the square root
distinguishes between the first and second turning points for both
the electron and the hole, and  $\nu_{\pm}=\nu_0 \pm 2 \varepsilon$.
Note that $\tau_{1}^{\pm} < \tau_{2}^{\pm}$, and the turning points
are real if either $m_{\pm}>0$ or $\nu_{\pm}\ge 2m_{\pm}$ for $m_{\pm}
<0$. For the electron and hole parts, the cyclotron radius
$\varrho_{\pm}$ and the guiding center $c_{\pm}$ are given~\cite{Lent}
by $\varrho_{\pm} =  l\, \sqrt{\nu \pm 2\varepsilon}$ and $c_{\pm}=l\,
\sqrt{\nu\pm 2\varepsilon + 2m_{\pm}}$, respectively.

\begin{table*}
\begin{tabular}{|c|p{23mm}|p{23mm}|p{23mm}|p{23mm}|p{23mm}|p{23mm}|}
\hline type of orbits & 
$A_1$ &  $A_2$ & $B_1$ & $B_2$ & $C_1$ & $C_2$ \\
& \includegraphics[scale=0.32]{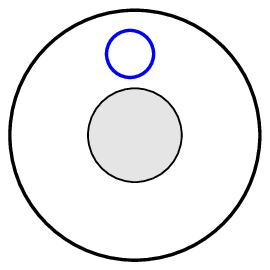} & 
\includegraphics[scale=0.32]{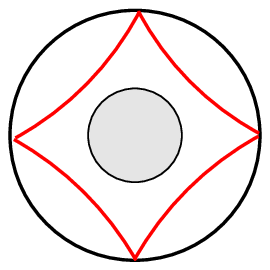} & 
\includegraphics[scale=0.32]{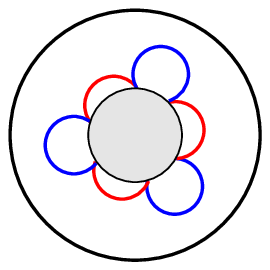} & 
\includegraphics[scale=0.32]{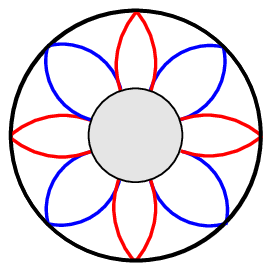} & 
\includegraphics[scale=0.32]{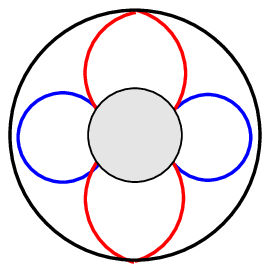} & 
\includegraphics[scale=0.32]{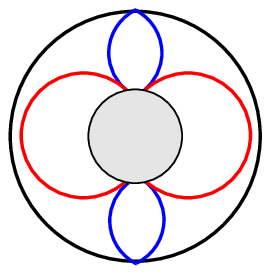} \\[2ex] \hline 
conditions &  
$\begin{matrix} \tau_S<\tau_1^{\pm} \cr \tau_2^{\pm}<\tau_N 
\end{matrix}$ & 
$\begin{matrix} \tau_S<\tau_1^{\pm} \cr \tau_N<\tau_2^{\pm}
\end{matrix}$ &
$\begin{matrix} \tau_1^{\pm} <\tau_S\cr \tau_2^{\pm}<\tau_N
\end{matrix}$ &
$\begin{matrix} \tau_1^{\pm} < \tau_S\cr\tau_N<\tau_2^{\pm}
\end{matrix}$ &
$\begin{matrix} \tau_1^{\pm} <\tau_S \cr\tau_2^{-}<\tau_N\cr\tau_N<
\tau_2^{+} \end{matrix}$ &
$\begin{matrix}\tau_1^{\pm} <\tau_S\cr\tau_2^{+}<\tau_N\cr \tau_N<
\tau_2^{-} \end{matrix}$   \\ \hline
\end{tabular}
\caption{Classification of orbits. The solid/dashed lines correspond
to the trajectory of an electron/hole. For the value of turning points
$\tau_{1,2}^\pm$, see Eq.~(\ref{turnpoints}).
\label{orbits:table}}
\end{table*}
The relative position of the turning points compared to $\tau_S=R_S/l$
and $\tau_N=R_N/l$ enables a classification of possible classical 
orbits which is summarized in Table \ref{orbits:table}. Orbits of
type $A_1$ correspond to the Landau states (or cyclotron orbits),
while $A_2$ are the so-called skipping orbits (or whispering--gallery
modes discussed, e.g., in Ref.~\onlinecite{Bruder}). In both cases,
the orbits do not touch the superconductor and, hence, electron and
hole states are not coupled. The other four types of orbits reach the
NS interface. In case of type $B_1$, electron and hole alternately
Andreev--reflect at the NS interface without ever touching the
boundary of the N region. For type $B_2$, the orbits reach both the
inner and the outer circles delimiting the N region. Finally, for
types $C_1$ and $C_2$, either the electron or the hole reaches the
outer circle.    
 
\begin{table}[b]
\begin{tabular}{|c|c|c|}\hline
 & $\Phi_m(\varepsilon)$ & $\mu$  \cr \hline \hline 
$A_1$ &
{\raisebox{-0.1cm}{$S_m^{\left(\pm\right)}(\tau_2^{\pm},
\tau_1^{\pm})$}} & $\frac{1}{2}$ \\ \hline
$A_2$ &
{\raisebox{-0.1cm}{$S_m^{\left(\pm\right)}(\tau_N,
\tau_1^{\pm})$}} & $\frac{3}{4}$ \\ \hline
 $B_1$ &
{\raisebox{-0.1cm}{$S_m^{\left(+\right)}(\tau_2^{+},\tau_S)
-S_m^{\left(-\right)}(\tau_2^{-},\tau_S)
-\frac{1}{\pi}\arccos \frac{E}{\Delta_0}$}} & 
$0$ \\ \hline
$B_2$ &
{\raisebox{-0.1cm}{$S_m^{\left(+\right)}(\tau_N,\tau_S)
-S_m^{\left(-\right)}(\tau_N,\tau_S)
-\frac{1}{\pi}\arccos \frac{E}{\Delta_0}$}} 
& $0$   \\ \hline
 $C_1$ &
{\raisebox{-0.1cm}{$S_m^{\left(+\right)}(\tau_N,\tau_S)
-S_m^{\left(-\right)}(\tau_2^{-},\tau_S)
-\frac{1}{\pi}\arccos \ \frac{E}{\Delta_0}$}} 
& $\frac{1}{4}$ \\ \hline
$C_2$ &
{\raisebox{-0.1cm}{$S_m^{\left(+\right)}(\tau_2^{+},\tau_S)
-S_m^{\left(-\right)}(\tau_N,\tau_S)
-\frac{1}{\pi}\arccos \frac{E}{\Delta_0}$}} 
&  $-\frac{1}{4}$\\ \hline 
\end{tabular}
\caption{Quantization conditions for the different orbits. See also
the text. 
\label{quant:table}}
\end{table}
In the S region, we approximate the wave function in the same way as 
in Ref.~\onlinecite{box_disk:cikk}. Substituting the corresponding WKB
wave functions and their derivatives into the secular equation
(\ref{DNS}) and assuming $R_S\gg \xi_0$, we obtain, after tedious but
straightforward algebra, the following quantization condition for the
semiclassically approximated energy levels:  
\begin{equation}
\Phi_m(\varepsilon) = n+\mu \quad .
\label{semi_DNS}
\end{equation}
Here  $n$ is an integer, and the phase $\Phi_m(\varepsilon)$ and 
the Maslov index $\mu$ are given in Table \ref{quant:table}. The
radial action $S_m^{\left(\pm\right)}$ (in units of $\hbar$) of the
electron and the hole between $\tau_1$ and $\tau_2$ reads 
\begin{subequations} 
\begin{equation}
S_m^{\left(\pm\right)}(\tau_2,\tau_1) = 
\Theta_m^{\left(\pm\right)}(\nu,\tau_2)
-\Theta_m^{\left(\pm\right)}(\nu,\tau_1) \quad , 
\label{S_m}
\end{equation}
where
\begin{eqnarray}
2\pi \Theta_m^{\left(\pm\right)}(\nu,\tau) &=& 
\sqrt{\tau^2\nu_{\pm}-\left(\frac{\tau^2}{2}-m_{\pm}\right)^2} 
\nonumber \\ 
& & \hspace*{-1.5cm}
-(\nu_{\pm}+m_{\pm})\arcsin\left(\frac{\nu_{\pm}+m_{\pm}
-\tau^2/2}{\sqrt{\nu_{\pm}(\nu_{\pm}
+2m_{\pm})}}\right)\nonumber \\
& & \hspace*{-1.5cm}
-|m_{\pm}|\arcsin\left(\frac{\tau^2(\nu_{\pm}+m_{\pm})
-2 m_{\pm}^2}{\tau^2\sqrt{\nu_{\pm}
(\nu_{\pm}+2m_{\pm})}}\right). \label{Theta}
\end{eqnarray}
\end{subequations}
Note that the value of the Maslov index comes out directly from our
semiclassical calculation. This result can be interpreted as follows. 
For orbits of type $A_1$, the quantization condition can be simplified
to  $\nu^{\pm}=n+\frac{1}{2}+\frac{1}{2}(m_{\pm}-|m_{\pm}|)$, which 
coincides with the quantization of the electron/hole cyclotron states
of a normal ring in a magnetic field. These are the familiar Landau
levels. For orbits of type $A_2$, the radial action between the
boundaries of the classical allowed region ($\tau_1^{\pm}$ and
$\tau_N$) is equal to $n+\mu$, where $\mu$ has a contribution 
$\frac{1}{4}$ from the soft turning point (at $\tau=\tau_1^{\pm}$),
and $\frac{1}{2}$ from the hard turning point (at $\tau=\tau_N$),
resulting in an overall $\mu=3/4$. For cases $B_1, B_2, C_1, C_2$,
Andreev reflection takes place at the NS interface, resulting in an
additional phase shift $-\frac{1}{\pi}\arccos \frac{E}{\Delta_0}$. The
action/Maslov index for the hole is $-1$ that of the action/Maslov
index of the electron between the same boundaries. The conditions for
the appropriate boundaries of these types of orbits can be obtained
from Table~\ref{orbits:table}. The Maslov index is zero for cases
$B_1$ and $B_2$ because the hole contribution cancels that of the
electron. At the outer boundary for type $C_1$, there is a hard
turning point for the electron and a soft turning point for the hole,
resulting in $\mu=\frac{1}{2}-\frac{1}{4}=\frac{1}{4}$. Similarly, for
$C_2$, we have  $\mu=-\frac{1}{2}+\frac{1}{4}=-\frac{1}{4}$.   

\begin{figure*}[hbt]
\includegraphics[width=5.8cm]{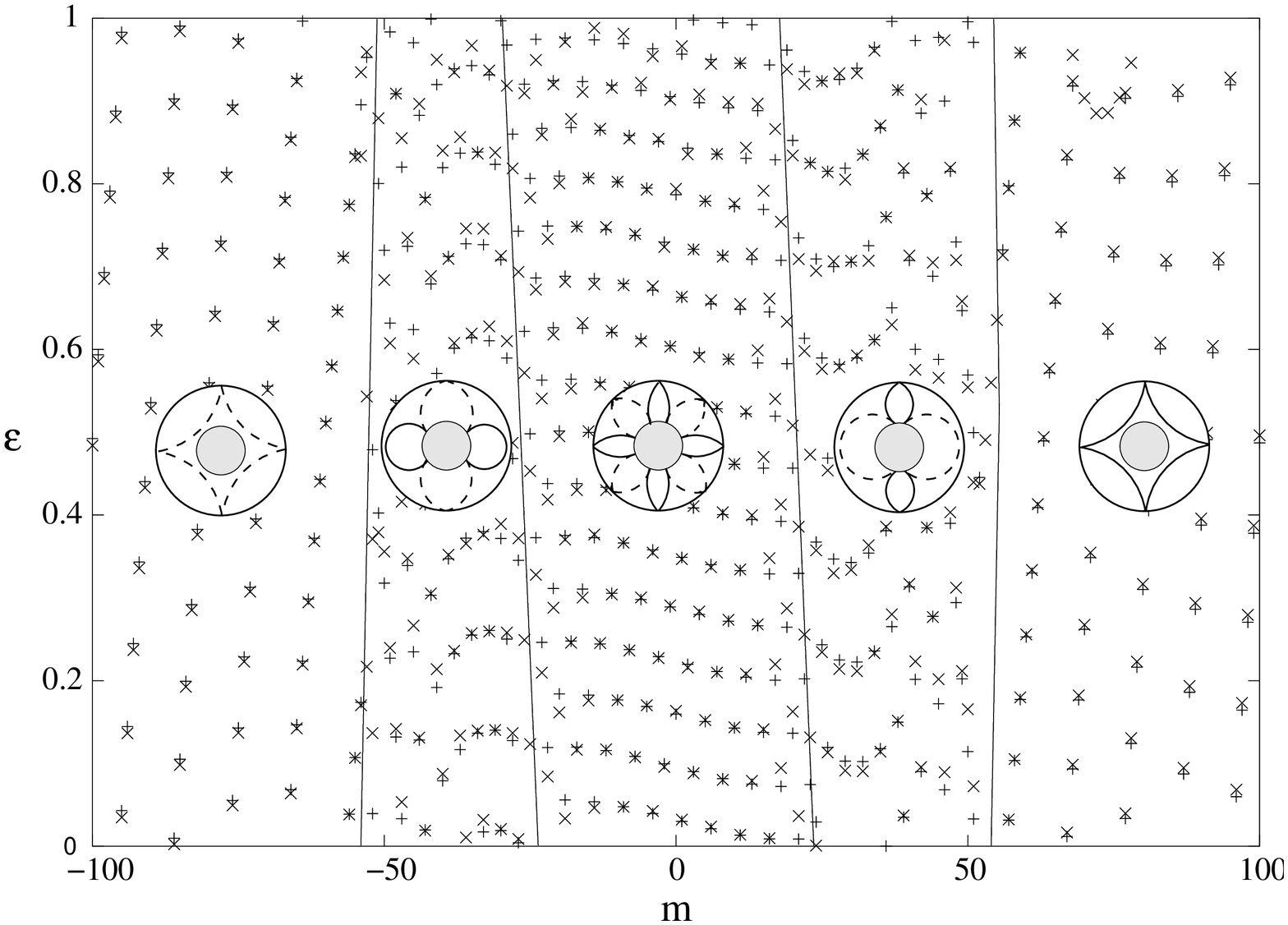}\hspace{1cm} 
\includegraphics[width=5.8cm]{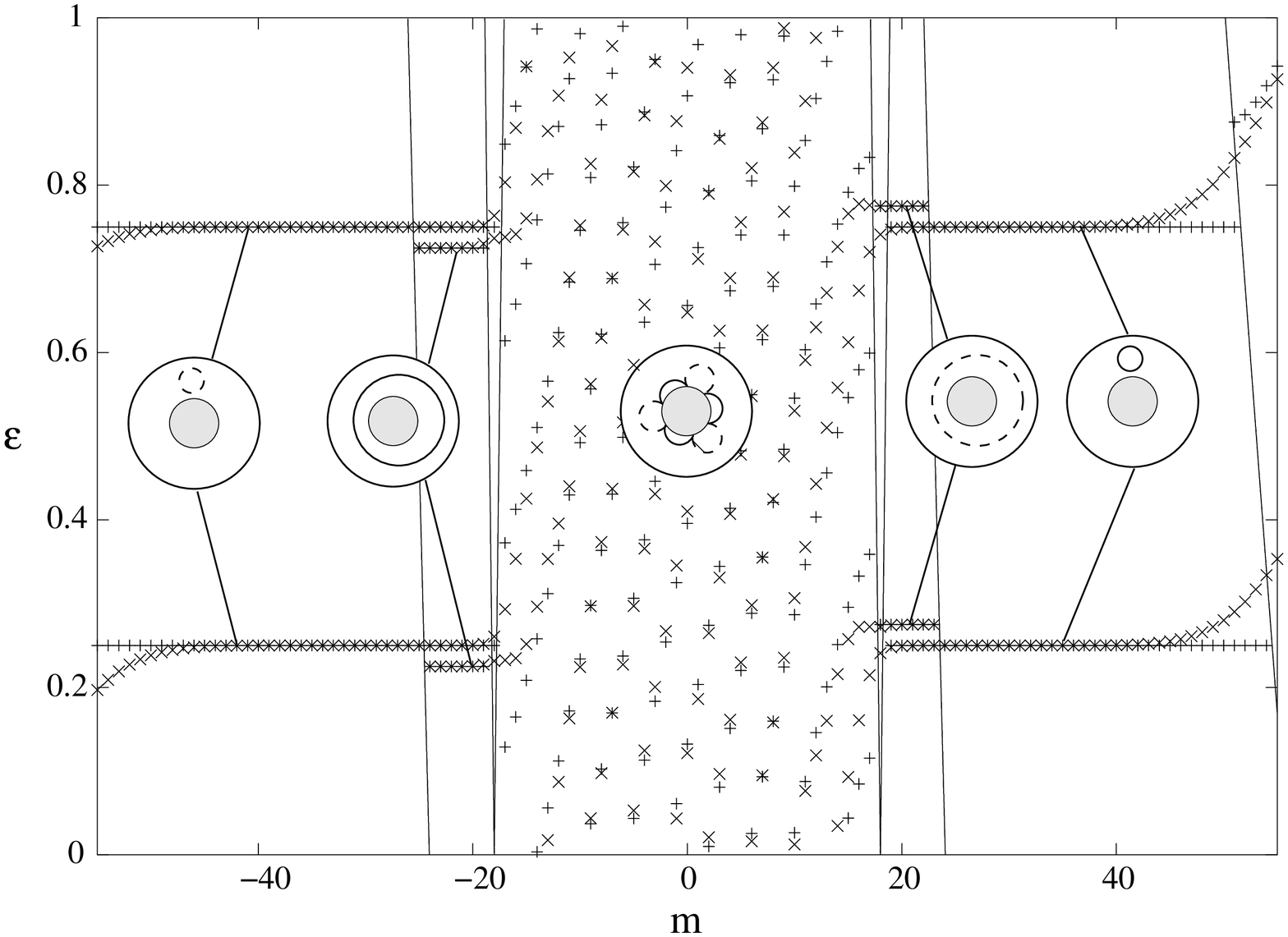}
\caption{Exact (crosses) and semiclassical (+ signs) energy levels (in
units of $\Delta_0$) obtained from Eqs.~(\ref{DNS}) and
(\ref{semi_DNS}) as functions of quantum number $m$. In the left/right
panel, $k_{\rm F}R_S=24.0, 18.0 $, $\Phi_{\rm miss}=7.2, 4.05$, and
$R_S/R_N=0.29981, 0.14996$, respectively. For both cases, $\Delta_0/
E_{\rm F}=0.1$. The solid lines represent the border of regions in the
$\varepsilon, m$ plane where the different types of orbits arise.
These lines can be obtained from conditions given in
Table~\ref{orbits:table}. For easy reference, the different types of
orbits are shown in the corresponding regions.
\label{Q-BS:fig}}
\end{figure*}

In numerical calculations, it is convenient to use the following
parameters that are suitable for characterizing the experimental
situation: $k_{\rm F}R_S$, where $k_{\rm F}$ is the Fermi wave number,
$\Phi_{\rm miss}=BR^2_S\pi$ is the missing flux due to the Meissner
effect, $R_S/R_N$ and $\Delta_0/E_{\rm F}$. Fig.~\ref{Q-BS:fig} shows
the comparison of the energy levels from the exact quantum calculation
with the semiclassical results for two different systems. One can see
that the semiclassical approximation is in excellent agreement with
the exact quantum calculations. In Fig.~\ref{Q-BS:fig}b, the magnetic
field is high enough for the appearance of Landau levels showing no
dispersion as function of $m$. A small difference between quantum and
semiclassical calculations occurs at the border of the region of the
(cyclotron) orbits $A_1$.

The experimental situation of Ref.~\onlinecite{Visani} corresponds to
the low-magnetic-field limit. There the cyclotron radius is large
compared to $R_N$, and only orbits of type $A_2$ and $B_2$ exist in
the semiclassical approximation. Therefore, only these two types  
contribute to the free energy and, ultimately, to the susceptibility.
In a simplified model, these orbits have been included in Bruder and
Imry's theoretical study~\cite{Bruder}. Thermodynamical quantities
such as the magnetic moment or the susceptibility can be determined
from the energy levels of the system~\cite{Carlo3}. However, to fully
explain the experimental results~\cite{Visani}, one needs to extend
the work presented in this paper. For example, the Meissner effect can
be included in a similar way as in Ref.~\onlinecite{Uli2}. The energy
levels above the bulk superconducting gap ($E > \Delta_0$) can be
obtained by analytical continuation of the secular equation
(\ref{DNS}). One can expect a negligible effect from the roughness of
the NS interface if the amplitude of the roughness is less than the 
wave length of the electrons~\cite{Klama1-2,disord}.


In conclusion, we presented a systematic treatment of an
experimentally relevant Andreev billiard in a magnetic field using
the Bogoliubov-de Gennes formalism. An exact secular equation for
Andreev--bound--state levels was derived that we evaluated both
numerically and using semiclassical methods. In particular, a
classification of possible classical electron and hole orbits in
arbitrary magnetic fields was presented. This provides a useful
starting point for successive studies of thermodynamic properties
because it is possible to obtain the free energy of an NS hybrid
system from the quasiparticle energy spectrum. Such an analysis may
shed further light on the origin of the recently observed paramagnetic
re-entrance effect and opens up a whole arena of new possibilities to
study Andreev billiards in magnetic fields.

One of us (J.\ Cs.) gratefully acknowledges very helpful discussions
with C.~W.~J.~Beenakker. This work is supported in part by the
European Community's Human Potential Programme under Contract No.
HPRN-CT-2000-00144, Nanoscale Dynamics, the Hungarian-British
Intergovernmental Agreement on Cooperation in Education, Culture,
Science and Technology, and the Hungarian  Science Foundation OTKA
TO34832.

\end{document}